\documentclass[conference]{IEEEtran}
\IEEEoverridecommandlockouts
\makeatletter
\renewcommand{\fnum@figure}{Fig. \thefigure}
\makeatother
\usepackage[ left=0.673in, right=0.673in, top=0.773in, bottom=1in]{geometry}
\usepackage[T1]{fontenc}  
\usepackage{comment}
\usepackage[noadjust]{cite}
\usepackage{filecontents}
\usepackage{balance}

\usepackage{graphicx}
\usepackage{epstopdf}
\usepackage{amsfonts}
\DeclareGraphicsExtensions{.pdf,.jpeg,.png,.eps}
\usepackage[cmex10]{amsmath}
\usepackage{mathabx}
\usepackage{algorithmic}
\usepackage{array}
\usepackage{mdwmath}
\usepackage{mdwtab}
\usepackage{float}
\usepackage{eqparbox}
\usepackage{url}
\usepackage{color,soul}
\usepackage{amsmath} 
\usepackage{algorithm} 
\usepackage{multirow} 

\usepackage{dirtytalk} 

\usepackage[utf8]{inputenc}
\usepackage[english]{babel}
\usepackage{fancyhdr}
\usepackage{lastpage}
\usepackage{caption}
\usepackage{subcaption}
\usepackage{adjustbox}
\usepackage{amssymb}
\usepackage{amsfonts}

\usepackage{xcolor}

\usepackage{calrsfs}
\DeclareMathAlphabet{\pazocal}{OMS}{zplm}{m}{n}

\usepackage[acronym]{glossaries}
\newacronym{IDS}{IDS}{intrusion detection systems}
\newacronym{CNN}{CNN}{convolutional neural networks}
\newacronym{LSTM}{LSTM}{long short-term memory}
\newacronym{AI}{AI}{artificial intelligence}
\newacronym{DL}{DL}{deep learning}
\newacronym{ML}{ML}{machine learning}
\newacronym{FP}{FP}{false positives}
\newacronym{FN}{FN}{false negatives}
\newacronym{TP}{TP}{true positives}
\newacronym{TN}{TN}{true negatives}
\newacronym{UAVs}{UAVs}{unmanned aerial vehicles}
\newacronym{UAV-IDS}{UAV-IDS}{unmanned aerial vehicle intrusion detection system}
\newacronym{PCA}{PCA} {principal component analysis}
\newacronym{Relu}{Relu}{rectified linear unit}
\newacronym{IoT}{IoT}{internet of things}
\newacronym{SIDS}{SIDS}{signature-based IDS}
\newacronym{AIDS}{AIDS}{anomaly-based IDS}
\newacronym{NIDS}{NIDS}{network intrusion detection system}
\newacronym{HIDS}{HIDS}{host-based IDS}
\newacronym{1D}{1D}{one dimensional}
\newacronym{RF}{RF}{radio frequency}
\newacronym{DNN}{DNN}{deep neural networks}
\newacronym{ViT}{ViT}{vision transformer}
\newacronym{CV}{CV}{computer vision}
\def\BibTeX{{\rm B\kern-.05em{\sc i\kern-.025em b}\kern-.08em
    T\kern-.1667em\lower.7ex\hbox{E}\kern-.125emX}}

\begin{document}

\bstctlcite{IEEEexample:BSTcontrol}
\title{A Conditional Tabular GAN-Enhanced Intrusion Detection System for Rare Attacks in IoT Networks
}

\author{\IEEEauthorblockN{Safaa Menssouri, and  El Mehdi Amhoud
}
\IEEEauthorblockA{College of computing, Mohammed VI Polytechnic University, Ben Guerir, Morocco 
} 
{\{safaa.menssouri, elmehdi.amhoud\}@um6p.ma}}

\maketitle

\begin{abstract}
Internet of things (IoT) networks, boosted by 6G technology, are transforming various industries. However, their widespread adoption introduces significant security risks, particularly in detecting rare but potentially damaging cyber-attacks. This makes the development of robust \gls{IDS} crucial for monitoring network traffic and ensuring their safety. Traditional \gls{IDS} often struggle with detecting rare attacks due to severe class imbalances in IoT data. In this paper, we propose a novel two-stage system called conditional tabular generative synthetic minority data generation with deep neural network (CTGSM-DNN). In the first stage, a conditional tabular generative adversarial network (CTGAN) is employed to generate synthetic data for rare attack classes. In the second stage, the SMOTEENN method is applied to improve dataset quality. The full study was conducted using the CSE-CIC-IDS2018 dataset, and we assessed the performance of the proposed IDS using different evaluation metrics. The experimental results demonstrated the effectiveness of the proposed multiclass classifier, achieving an overall accuracy of 99.90\% and 80\% accuracy in detecting rare attacks.
\end{abstract}

\section{Introduction}
The proliferation of \gls{IoT} networks, coupled with the emergence of 6G technology \cite{9598915}, are transforming numerous aspects of modern life, from smart homes and cities to industrial automation and healthcare systems. \gls{IoT} networks comprise a vast array of interconnected devices that collect, exchange, and analyze data to facilitate intelligent decision-making and automation. This interconnected ecosystem enhances efficiency, productivity, and convenience, driving the rapid adoption of \gls{IoT} technologies across various domains. However, the widespread deployment of \gls{IoT} devices within the high-speed, low-latency framework of 6G introduces substantial security challenges. The heterogeneous nature of these networks, combined with their large-scale and distributed architecture, makes them vulnerable to a variety of cyber threats \cite{7562568}. Traditional security measures are often inadequate for addressing the unique requirements and vulnerabilities inherent in \gls{IoT} environments. Thus, robust and adaptive security solutions, such as intrusion detection systems (IDS) \cite{9984320}, are essential for monitoring network traffic, detecting anomalies, and safeguarding against potential security breaches.

Although there are different types of IDS, in our research work we are interested in anomaly-based IDS (AIDS) \cite{9620099}. These systems aim to detect both known and unknown attacks, providing a more comprehensive detection capability than other IDS.\\
One of the significant challenges in developing effective anomaly-based IDS is the issue of imbalanced data. In many real-world scenarios, the number of normal instances vastly outweighs the number of anomalous ones, creating a class imbalance problem. This imbalance is particularly problematic when dealing with rare classes of attacks.

Rare attacks are a type of security breach that occur infrequently and have a low number of instances, making them less familiar and more challenging to detect. 
They often account for only a small fraction of the normal data, and when using a single \gls{ML} model for classification, the classifier tends to favor the majority class of data and misclassify the rare-class attack data.\\
Despite their rarity, these attacks can indicate more sophisticated and potentially damaging threats. As a result, developing effective techniques to detect rare intrusion events is crucial for maintaining the security and integrity of \gls{IoT} networks.
In this work, we aim at enhancing the detection capabilities for rare attacks by exploring novel approaches and leveraging advanced \gls{ML} techniques. While several previous studies have focused on using generative adversarial networks (GANs) for data augmentation to address the challenge of imbalanced datasets, our approach is distinguished by the use of conditional tabular GAN (CTGAN). Unlike traditional GANs, which can struggle with discrete data characteristics, CTGAN excels at capturing the nuances of rare attack classes. \\
The main contributions of this paper are summarized as follows:
\begin{itemize}
    \item We develop a novel two-stage methodology that first employs CTGAN to generate synthetic data for rare attack classes, followed by the SMOTEENN method to enhance the detection of attacks in IoT networks.
    \item  We validate our proposed system on one of the largest public datasets (CSE-CIC-IDS2018) for IDS, and we conduct comprehensive comparisons with existing methods.
    \item The experiments demonstrate that our proposed solution significantly improves the detection accuracy of rare attack instances, achieving up to 80\% accuracy for rare attacks while maintaining 99.90\% overall classification accuracy.
\end{itemize}
The remainder of the paper is organized as follows:\\ In section II, we introduce the related work. In section III, we detail the architecture of our proposed model along with a description of the dataset. In section IV, we present our simulation results, and we discuss the observations and findings. Finally, in section V, we conclude and outline our perspectives.

\section{Related work}
Intrusion detection systems play a crucial role in safeguarding \gls{IoT} networks. Numerous studies have focused on developing effective \gls{IDS} for \gls{IoT} networks, employing a variety of techniques ranging from traditional signature-based methods to more advanced \gls{ML} approaches. Notably, supervised learning algorithms, such as k-nearest neighbors (KNN), naïve bayes, and support vector machines (SVM) \cite{9994392}, have been extensively studied for their effectiveness in detecting anomalies. In \cite{8080566}, the authors conducted  several experiments to evaluate the efficiency and performance of various \gls{ML} classifiers, such as random forest, random tree, decision table, naive Bayes, and Bayes network. All the tests were conducted using the KDD dataset. The experiments demonstrated that there is no single \gls{ML} model that can handle efficiently all the types of attacks.
Furthermore, \gls{DL} techniques, such as \gls{CNN} or \gls{LSTM} \cite{9817245} are widely used in \gls{IDS}.\\
\indent The use of a single \gls{ML} model has inherent \mbox{limitations \cite{10419091}.} Thus, in recent years, various learning algorithms have been combined to enhance performance of \gls{IDS} \cite{menssouri}. For instance, \mbox{in \cite{10064274}} the authors proposed an \gls{IDS} that combines the powerful learning ability of LSTM in time series data with \gls{CNN}'s ability to extract spatial features. The model was trained using KDD CUP99, NSL-KDD, and UNSW-NB15 classic datasets, the results show good convergence and performance.\\
\indent Despite these advancements, a significant challenge remains in dealing with imbalanced datasets, particularly regarding rare attacks detection. Traditional and even some advanced \gls{ML} models often struggle to accurately detect rare attacks due to their scarcity in the training data. Xu et al. \cite{8449272} presented a recurrent neural network based \gls{IDS}. Their approach, tested on the NSL-KDD and KDD Cup'99 datasets, demonstrated promising results compared to other methodologies. However, it has a limitation in detecting minority attack classes, such as U2R and R2L, resulting in lower detection rates for these specific classes. Other works, such as \cite{9172014}, have utilized ensemble learning with feature selection technique to enhance IDS performance on the CSE-CIC-IDS2018 dataset. While the overall detection accuracy has been improved, rare attack classes like SQL Injection and Brute Force were not well classified. Additionally, the infiltration attack class showed lower performance, revealing challenges in detecting these types of attacks. Moreover, in \cite{10044208}  the authors proposed a Bagging-DNN-based \gls{IDS} that addresses class imbalance by incorporating class weights and leveraging \gls{DNN} as core estimators. Their method achieved high performance across four datasets, with accuracy reaching 98.90\%, demonstrating effective detection of both known and unknown attacks. However, their experiments were limited to binary classification, which may restrict its applicability in multiclass scenarios, particularly for detecting rare attacks.

Existing works address the issue of imbalanced data by employing class-weighting techniques, undersampling, or oversampling. Class weighting effectively enhances rare attack detection without introducing redundancy, but it may lead to trade-offs in majority class performance and struggles with extremely rare classes. On the other hand, undersampling can lead to the loss of valuable information, while oversampling can introduce redundancy and overfitting. Alternatively, some approaches employ ensemble models to improve rare attack detection, although these methods can be computationally complex, especially when dealing with large dataset.

 \begin{figure}[t]
\centering
\includegraphics[width=0.41\textwidth]{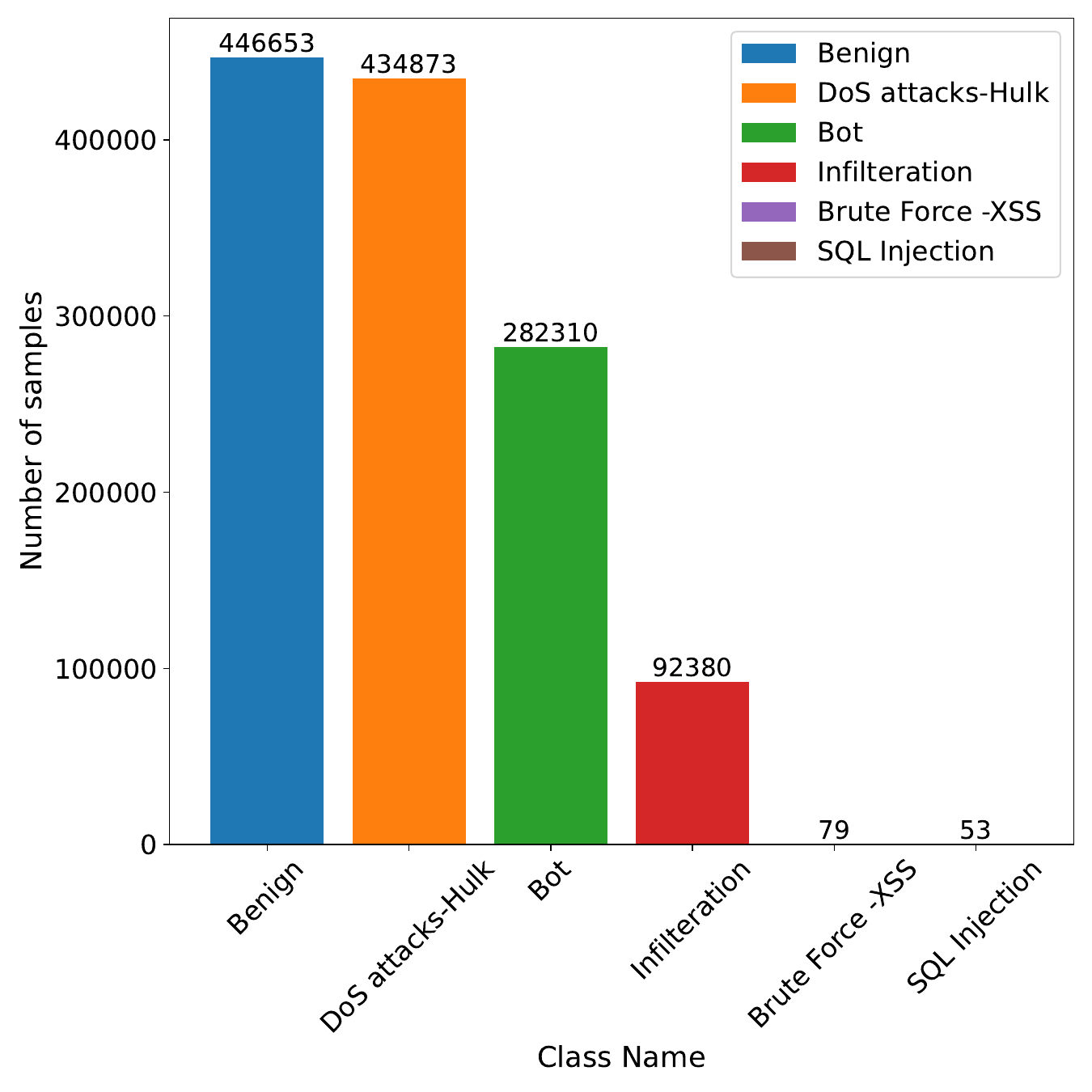}
    \caption{Class distribution of the dataset.}
    \label{fig: DataDistribution}
\end{figure}

\begin{table}[t]
  \caption{Selected files from the CSE-CIC-IDS2018 dataset}
  \resizebox{0.48\textwidth}{!}{
  \centering
    \begin{tabular}{c|c}
    \hline 
       \textbf{File name} &   \textbf{Class types}\\
       \hline \\
       Friday-16-02-2018 TrafficForML CICFlowMeter.csv  & Benign\\ & DoS attacks–Hulk \\[0.5em]

       Thursday-22-02-2018 TrafficForML CICFlowMeter.csv
       & Brute Force–XSS\\[0.5em]
       
       Friday-23-02-2018 TrafficForML CICFlowMeter.csv
       & SQL Injection\\[0.5em]
       
       Thursday-01-03-2018 TrafficForML CICFlowMeter.csv
       & Infilteration\\[0.5em]
       
       Friday-02-03-2018 TrafficForML CICFlowMeter.csv & Bot\\[0.5em]
       \hline                  
    \end{tabular}
    }
    \label{tab: Selected datasets}
\end{table}

\section{Methodology}
\subsection{CSE-CIC-IDS2018 Dataset}
In our study, we employed the open-source CSE-CIC-IDS2018 dataset, created by the Canadian Institute for Cybersecurity \cite{Sharafaldin2018}. This dataset was selected for its comprehensive and up-to-date nature, meeting essential criteria such as extensive traffic data, a variety of attack types, and detailed labeling. It encompasses seven distinct attack scenarios: Brute Force, Heartbleed, Botnet, DoS, DDoS, Web attacks, and network infilteration, as well as the benign data. Importantly, the CSE-CIC-IDS2018 dataset exhibits imbalanced classes, with certain rare attack types such as Brute Force-XSS and SQL Injection, occurring less frequently than others. The distribution of the dataset used in our analysis is illustrated in Fig. \ref{fig: DataDistribution}. From the figure, we notice that the number of samples of Brute Force-XSS and SQL Injection represent only 0.009\% and 0.006\%, respectively, from the total number of samples of all attacks. This rarity is crucial to our objective of developing an \gls{IDS} tailored for identifying rare attacks in \gls{IoT} networks.\\
In addition, the dataset contains 80 features extracted from network traffic and system logs. Although it comprises several files, only a selected subset was used for this study based on their relevance to the analysis. The chosen files and their corresponding class types are detailed in Table \ref{tab: Selected datasets}. The preprocessing step included merging these files, and the workflow is outlined in the following subsections.

\begin{algorithm}
\caption{SMOTEENN method}
\label{alg:smoteen}
\begin{algorithmic}[1]
\STATE \textbf{Input:} Dataset $D$ with minority class samples $D_{min}$ and majority class samples $D_{maj}$, Number of synthetic samples $N$, Number of nearest neighbors $k$
\STATE \textbf{Output:} Resampled set $D'$
\STATE $S \gets \emptyset$

\FOR{each $x_i \in D_{min}$}
    \STATE $N_i \gets k$-nearest neighbors of $x_i$ from $D_{min}$
\ENDFOR

\FOR{each $x_i \in D_{min}$}
    \FOR{$j \gets 1$ to $N$}
        \STATE Select $x_{ij} \in N_i$
        \STATE $x_{new} \gets x_i + \lambda (x_{ij} - x_i) $ where $ \lambda \sim U(0,1)$ 
        \STATE $S \gets S \cup \{x_{new}\}$
    \ENDFOR
\ENDFOR

\STATE $D' \gets D \cup S$

\FOR{each $x_i \in D'$}
    \STATE $N_i \gets k$-nearest neighbors of $x_i$
    \IF{$y_i \neq \text{majority class in } N_i$}
        \STATE $D' \gets D' \setminus \{x_i\}$
    \ENDIF
\ENDFOR
\end{algorithmic}
\end{algorithm}

\subsection{SMOTEENN Method}
SMOTEENN is a hybrid technique that combines synthetic minority over-sampling technique (SMOTE) \cite{10.5555/1622407.1622416} and edited nearest neighbors technique (ENN) to tackle imbalanced datasets \cite{10.1145/1007730.1007735}. Initially, SMOTE generates synthetic samples of the minority class to balance the dataset. Following this, the ENN method is applied to cleanse the data by removing noisy and ambiguous samples. This dual approach not only enhances the representation of rare attacks but also improves dataset quality by eliminating misleading instances, thereby enhancing the accuracy and reliability of intrusion detection. The SMOTEENN method is formalized in Algorithm \ref{alg:smoteen}.
\subsection{Conditional Tabular GAN (CTGAN)}
CTGAN is a specialized GAN designed for realistic tabular data generation \cite{NEURIPS2019_254ed7d2}, adept at handling mixed data types and complex dependencies. It employs conditional modeling,  training-by-sampling techniques, and mode-specific normalization to optimize synthesis across categorical and continuous variables, thereby enhancing data augmentation and effectively addressing class imbalance. CTGAN's adversarial training framework ensures continual improvement in generating realistic samples, making it ideal for various applications needing precise synthetic data from tabular datasets. In CTGAN, the generator \( G \) and discriminator \( D \) are trained simultaneously, with the goal of \( G \) producing synthetic data that \( D \) cannot distinguish from real data.

\begin{raggedright}
CTGAN employs mode-specific normalization using a Gaussian mixture model (GMM) for continuous variables and one-hot encoding for categorical variables. The overall probability density function $ p(x) $ of a continuous variable $ x $ is given by: 
\end{raggedright}
\begin{equation}
p(x) = \sum_{k=1}^{K} \pi_k \mathcal{N}(x \mid \mu_k, \sigma_k^2),
\end{equation}
where \( p(x) \) is formed by a weighted sum of \( K \) Gaussian components. Each component is characterized by its mean \( \mu_k \), variance \( \sigma_k^2 \), and weight \( \pi_k \), which determines its contribution to the mixture model. This ensures that the synthetic data preserves the statistical properties of the original data. 
\begin{figure*}[t]
\vspace{-0.5em}
\centering
\includegraphics[width=0.8\textwidth]{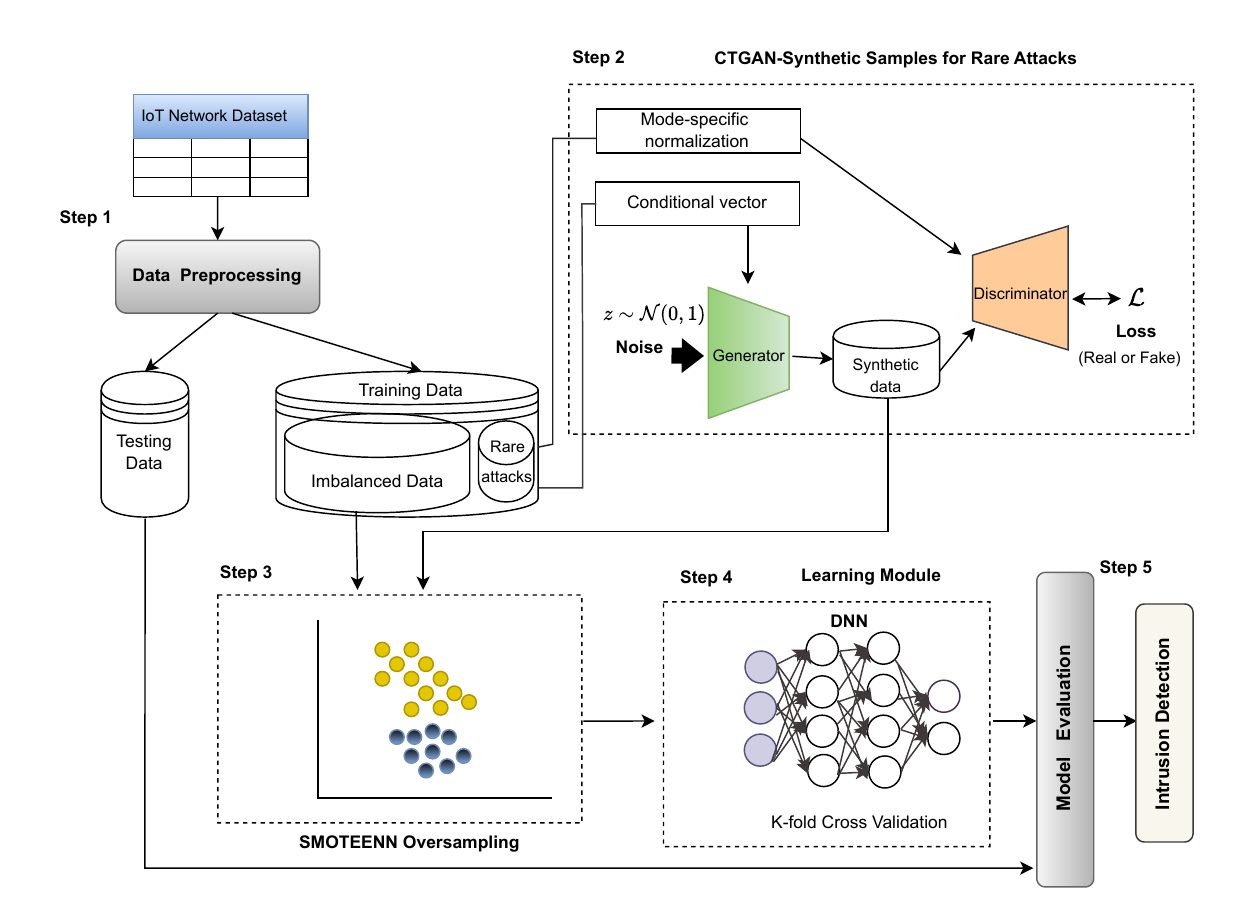}
    \caption{The proposed CTGSM-DNN intrusion detection system architecture.}
    \label{fig: CTGSM Model}
    \vspace{-0.5em}
\end{figure*}
\subsection{CTGSM-DNN System}
\begin{figure*}[t]
\centering
     \begin{subfigure}[b]{0.325\textwidth}
         \includegraphics[width=\textwidth]{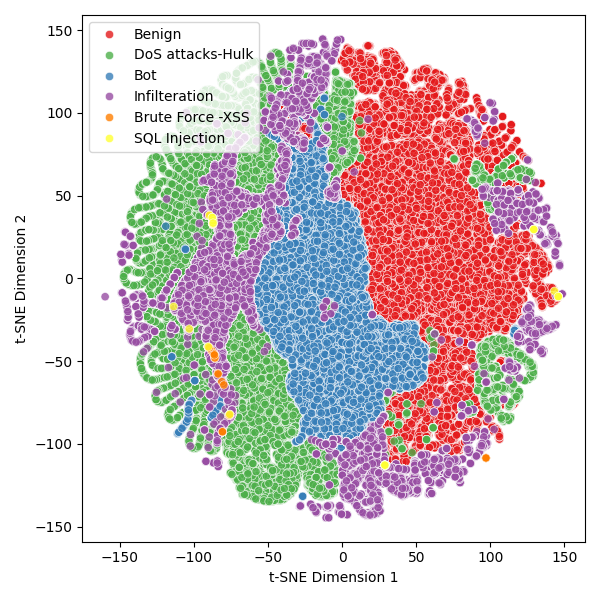}
         \caption{Original training data.}
         \label{fig: tsne_originalData}
     \end{subfigure}
          \begin{subfigure}[b]{0.335\textwidth}
         \includegraphics[width=\textwidth]{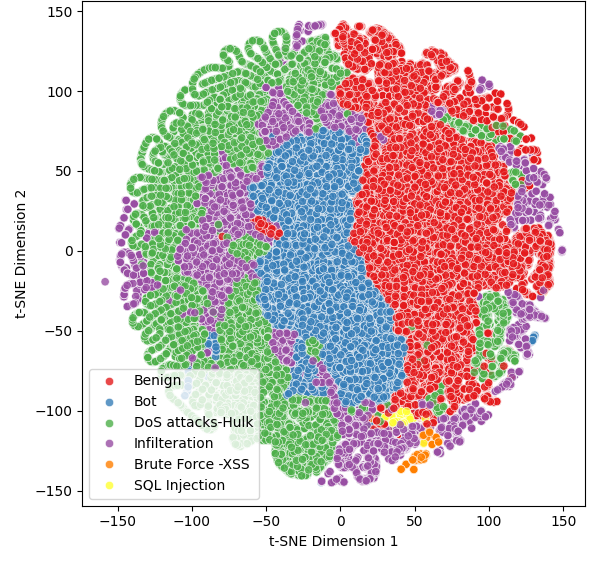}
         \caption{Training data after applying CTGAN.}
         \label{fig: tsne_Data_CTGAN}
     \end{subfigure}
     \begin{subfigure}[b]{0.325\textwidth}
     \includegraphics[width=\textwidth
        ]{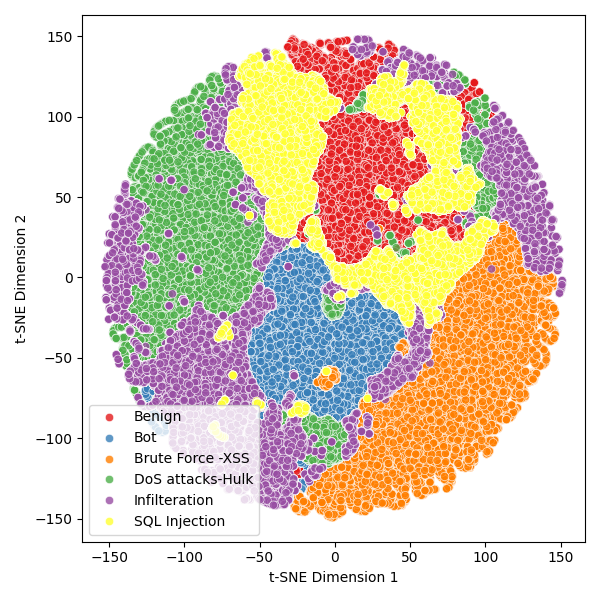}
         \caption{Training data after applying CTGAN and SMOTEENN.}
         \label{fig: tsne_Data_CTGSM}
     \end{subfigure}
        \caption{T-SNE visualization of CSE-CIC-IDS2018 training data before and after applying CTGAN and SMOTEENN methods.}
        
        \label{fig:TSNE_SMOTE_CTGAN}
\end{figure*}

The proposed conditional tabular generative synthetic minority data generation with deep neural network (CTGSM-DNN) intrusion detection system  consists of several sequential steps, as depicted in Fig. \ref{fig: CTGSM Model}. We began by importing the CSE-CIC-IDS2018 dataset and performing essential preprocessing tasks, including data cleaning, encoding, and normalization. As part of these steps, we also dropped the $timestamp$ feature from the dataset. To address the rarity of certain attacks, we employed CTGAN model to generate synthetic samples for the Brute Force-XSS and SQL Injection classes. This is shown in step 2 in Fig. \ref{fig: CTGSM Model}. This approach ensures the generation of realistic and less biased synthetic data, enriching the dataset with these rare attack instances. By incorporating these synthetic samples into the existing data, we applied the SMOTEENN method to balance the dataset, ensuring a cleaner and more representative distribution. (this is shown in step 3 in Fig. \ref{fig: CTGSM Model}). This combination of CTGAN and SMOTEENN is motivated by the need to balance augmentation with data quality. While CTGAN is effective at generating realistic synthetic samples, excessive augmentation (e.g., increasing from 30 to 100,000) can introduce redundancy or unrealistic samples, potentially leading to model overfitting or degraded performance. SMOTEENN addresses this issue by refining the dataset, to ensure the generated data is high-quality and representative, thereby enhancing training robustness.\\
Afterwards, in step 4, the resulting dataset is used to train the \gls{DNN} model. \\
\begin{table}[t]
  \caption{Deep neural network architecture and configuration details}
    \centering
    \begin{tabular}{c|c|c}
    \hline
       \textbf{Model} & \textbf{Criteria} &  \textbf{Values}\\
       \hline
                   CTGAN & Epoch & 700\\
                       & No. of synthetic samples & 1000 \\
        \hline
                       &Model & Sequential\\
                       & No. of hidden layers & 3\\
                       & Size of Input & 78\\
                       & No. of neurons in hidden layers & 128, 64, 6\\
                  DNN & Optimizer & Adam\\
                       & Activation function for hidden layer & ReLu\\
                       & Activation function for output layer & Softmax\\
                       & Dropout technique & Standard (p=0.4)\\
                       & Epoch / Batch\_size & 30 / 512\\   
       \hline
    \end{tabular}
    \vspace{-0.5em}
    \label{tab: Parameters of the System}
\end{table}
The impact of the CTGAN and SMOTEENN methods on the dataset distribution is visually demonstrated in Fig. \ref{fig:TSNE_SMOTE_CTGAN} using t-distributed stochastic neighbor embedding (t-SNE), a dimensionality reduction technique that reduces high-dimensional data to two dimensions while preserving local structures. \mbox{Fig. \ref{fig: tsne_originalData}} represents the original training data, highlighting the inherent class imbalance. In Fig. \ref{fig: tsne_Data_CTGAN}, the training data after applying CTGAN is shown, revealing the generation \mbox{of synthetic} samples for rare attacks. Finally, Fig. \ref{fig: tsne_Data_CTGSM} depicts the training data after applying both the CTGAN and SMOTEENN methods, illustrating a more balanced distribution among classes.\\
We trained and validated our model using 70\% of the dataset with a stratified five-fold cross-validation approach, which helps in achieving robust and reliable performance.\\
The effectiveness of the proposed model was evaluated on the remaining 30\% of the dataset using various metrics, including accuracy, loss, and confusion matrix. These metrics provide a thorough evaluation of the model's ability to detect rare and imbalanced classes, which is essential for improving IDS.
\subsection{Performance Evaluation Metrics}
To assess the performance of our proposed \gls{IDS}, we used various metrics, including accuracy ($AC$), precision ($PR$), recall ($R$), and F1-score ($F1$) via confusion matrix, along with the loss and the receiver operating characteristic (ROC) curve. These performance metrics are defined as follows:
\begin{equation}
AC=\frac{TP+TN}{TP+TN+FP+FN}, \mathrm{~}  R=\frac{TP}{TP+FN},
\end{equation}
\begin{equation}
F1=\frac{2TP}{2TP+FP+FN}, \mathrm{~}  PR=\frac{TP}{TP+FP}.
\end{equation}

With $TP$, $TN$, $FP$ and $FN$ representing true positives, true negatives, false positives, and false negatives, respectively.\\
\indent To train our model, we employed the focal loss, a cost-sensitive loss function designed to address class imbalance by modifying the cross-entropy loss function. It focuses on improving the learning of minority class samples.\\  Mathematically, the focal loss ($FL$) is expressed as:
\begin{equation}
FL = - \alpha (1 - p)^\gamma \log(p),
\end{equation}
where $p$ is the predicted probability of a data instance, $\alpha$ is a hyperparameter that balances the importance of positive and negative samples, and $\gamma$ is a focusing parameter that denotes a prefixed positive scale value.
\begin{figure}[t] 
\centering
\includegraphics[scale=0.4]{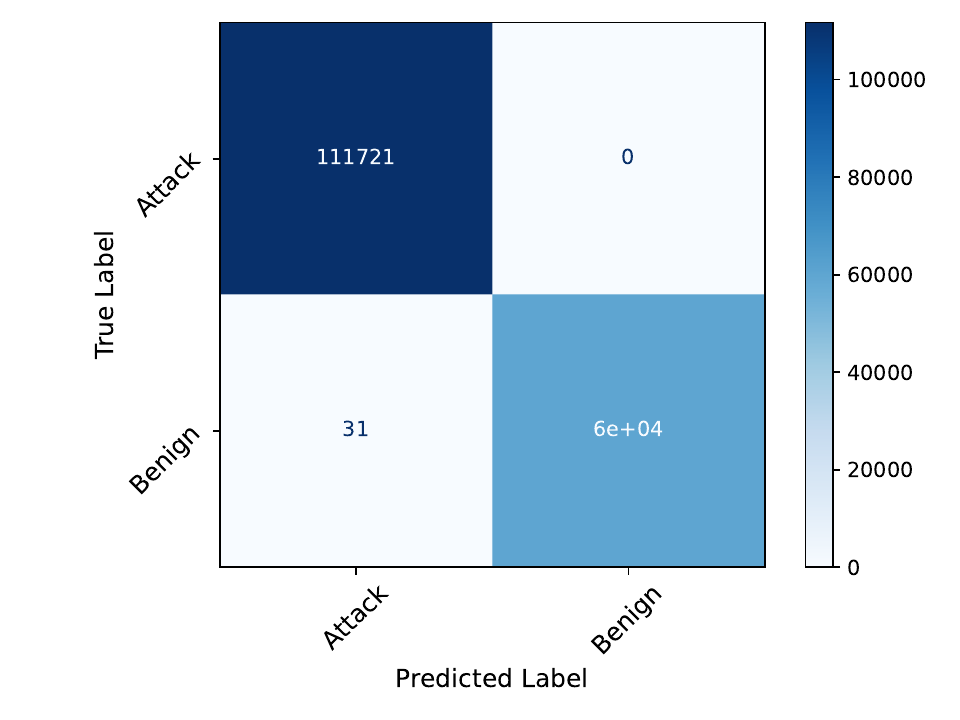}
     \caption{ Confusion matrix for binary classification.}
        \label{fig: ConfusionMatrixB}
\end{figure}
\section{Experimental Results}
In this section, we evaluate the performance of our proposed CTGSM-DNN system for detecting rare attacks. To ensure optimal model performance, we tuned the hyperparameters to find the most suitable ones, as presented in Table \ref{tab: Parameters of the System}. 
\subsection{Binary Classification}
In our experiments, we conducted the binary classification based on two labels of the data: benign and attack. \mbox{Fig. \ref{fig: ConfusionMatrixB}} illustrates the confusion matrix for the binary classifier model. The obtained results show the high performance of our proposed model in differentiating between normal and attack classes when applied to unseen data, with almost 99.98\% accuracy. Furthermore, our model maintains an impressive 99.99\% performance when considering other evaluation metrics, such as F1-score, recall, and precision, as shown in Table \ref{tab: Performance metrics}.
\subsection{Multiclass Classification}
\begin{figure}[t] 
\vspace{-0.5em}
\hspace{-1em}
\includegraphics[scale=0.3]
{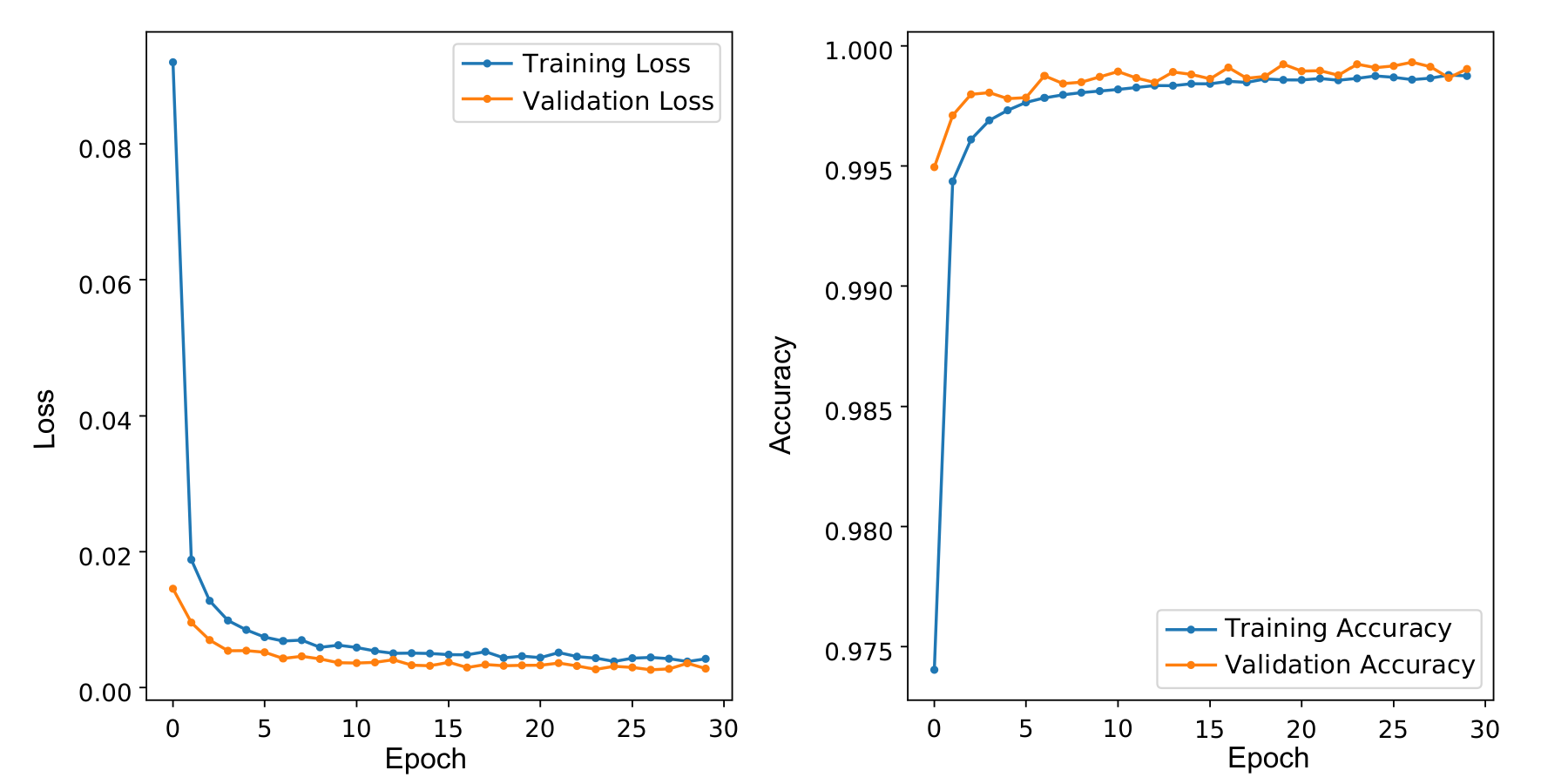}
        \caption{ Train accuracy and loss for multiclass classification.}
        \label{fig: TrainAccLossM}
\end{figure}

\begin{figure}[t] 
\centering
\hspace{-3.9em}
\includegraphics[scale=0.4]{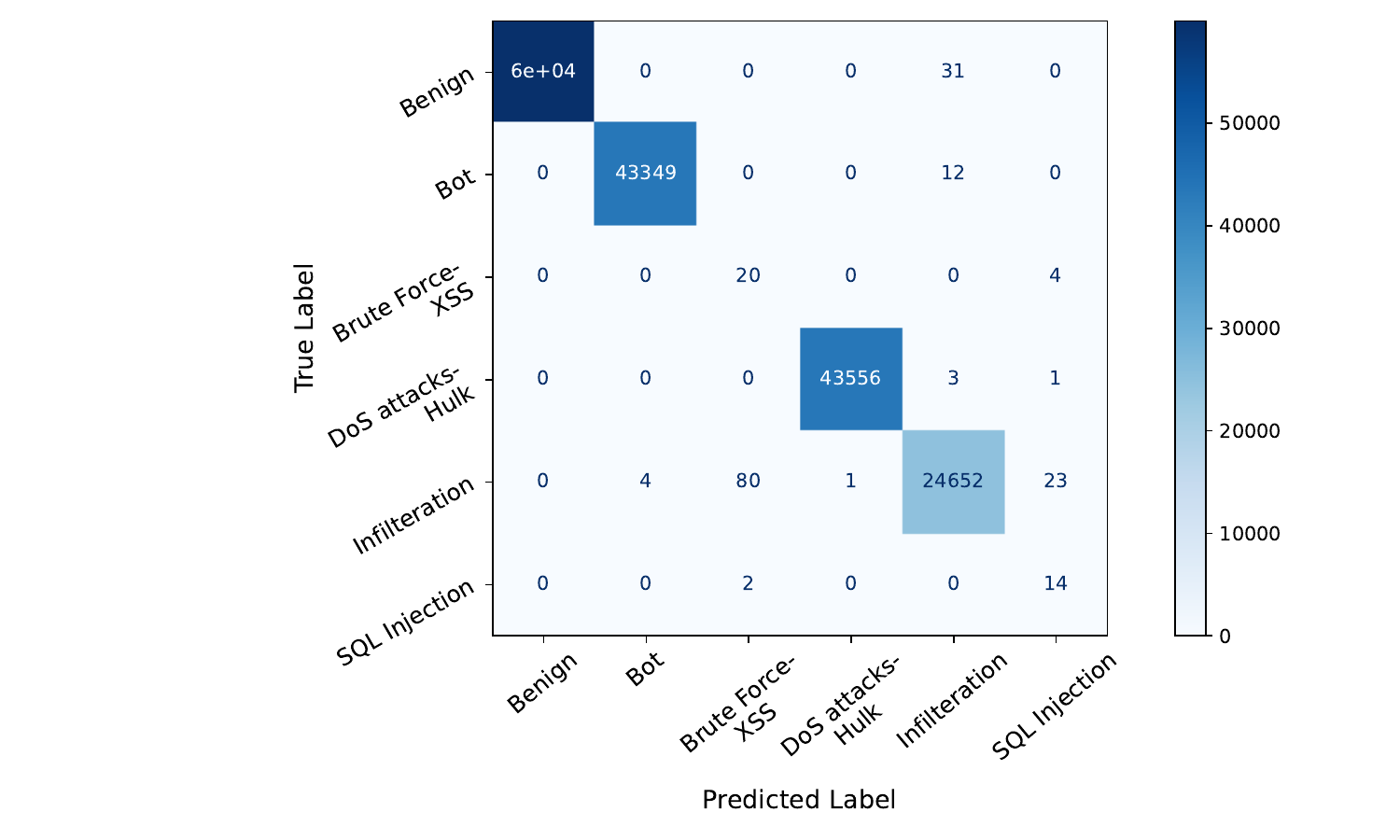}
     \caption{ Confusion matrix for multiclass classification.}
        \label{fig: ConfusionMatrixM}
        \vspace{-0.5em}
\end{figure}

\begin{figure}[t]  
\centering
\includegraphics[scale=0.4]{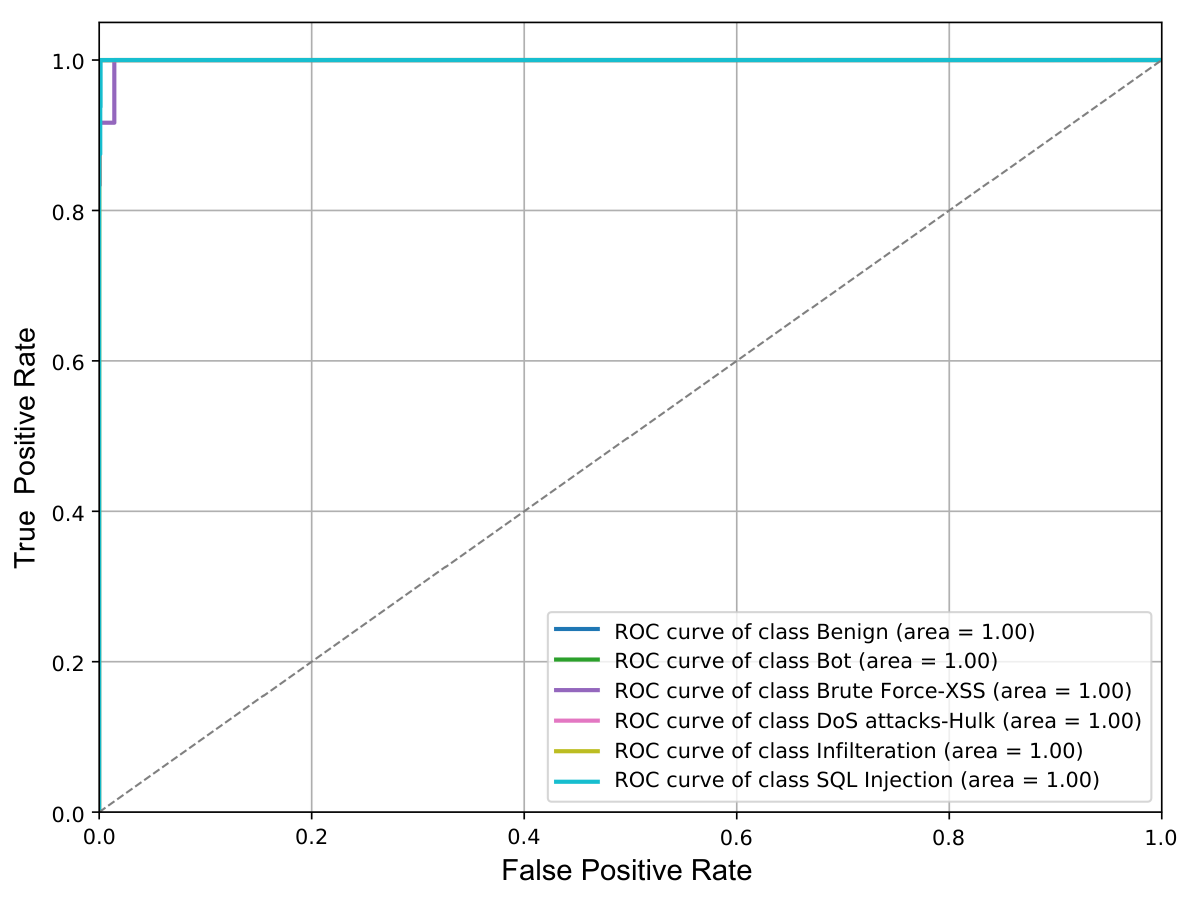}
     \caption{ Roc curve for multiclass classification.}
        \label{fig: RocCurveM}
        \vspace{-0.5em}
\end{figure}

Since our objective is to identify rare attacks, we conducted a multiclass classification task, where multiple attack types were classified alongside benign case. The experiments demonstrate that our proposed solution significantly improves the detection accuracy of rare attack instances, achieving up to 80\% accuracy for rare attacks while maintaining an overall classification accuracy of 99.90\%. Additionally, the model reached 99.90\% accuracy after only few epochs, with the loss converging to nearly 0\%, as shown in Fig. \ref{fig: TrainAccLossM}. Moreover, the confusion matrix presented in Fig. \ref{fig: ConfusionMatrixM}, highlights the model's ability to correctly classify most attack classes, with minimal misclassifications. Notably, for SQL Injection, 14 attacks were perfectly detected among 16, and for Brute Force-XSS, \mbox{20 attacks} were detected among 24. We also compared our model with other state-of-the-art methods, and the results as depicted in Table \ref{tab: Accuracy comparaison} show that our proposed approach achieved the highest recall and F1-score, further highlighting its efficacy. Finally, the ROC curves for each class, as shown in Fig. \ref{fig: RocCurveM}, demonstrate strong performance, with a true positive rate of approximately 99\% and a false positive rate of \mbox{around 1\%.} These results highlight our model's capability in detecting rare attacks.
\begin{table}[t]
 \centering
  \caption{Performance metrics (\%) of the binary classifier}
    \centering
    \begin{tabular}{|c|c|c|c|}
    \hline
    \centering
    \textbf{Class} & \textbf{Precision
} &  \textbf{Recall} &  \textbf{F1-score}\\
       \hline
       \centering
         Benign & 99.99 & 99.99 & 99.99 \\
        \hline
        \centering
        Attack & 99.99 & 99.99 & 99.99 \\ 
       \hline      
    \end{tabular}
    \label{tab: Performance metrics}
\end{table}

\begin{table}[t]
 \centering
  \caption{Averaged evaluation of methods on the CSE-CIC-IDS2018 dataset}
    \centering
    \begin{tabular}{|p{3.7cm}|c|c|c|}
    \hline
    \centering
    \textbf{Model} & \textbf{Precision
} &  \textbf{Recall} &  \textbf{F1-score}\\
       \hline
       \centering
        DNN \cite{7881684}& \textbf{1} & 0.75 & 0.79  \\
        \hline
        \centering
        SVM (with class weights) & 0.69 & 0.92 & 0.71 \\ 
       \hline
        \centering
        DNN (with class weights) \cite{9324926} & 0.67 & 0.94 & 0.68 \\ 
       \hline
        \centering
        DNN+SMOTE & 0.73 & 0.93 & 0.77 \\ 
       \hline
       \centering
        CNN+SMOTEENN \cite{8962490} & 0.70 & 0.94 & 0.72  \\
        \hline
        \centering
       CTGSM-DNN (Categorical Cross-Entropy loss)& 0.75 & 0.94 & 0.80 \\
       \hline
       \centering
       \textbf{Proposed Approach}& 0.80 & \textbf{0.95} & \textbf{0.84}\\
       \hline      
    \end{tabular}
    \label{tab: Accuracy comparaison}
\end{table}
\section{Conclusion}
In this paper, we proposed a new \gls{IDS} designed to tackle the challenge of detecting rare attacks in \gls{IoT} networks, which are often hindered by severe class imbalances in the data. By leveraging conditional tabular GAN for synthetic rare data generation and the SMOTEENN method for effectively tackling class imbalance and refining data quality, our approach enhances the detection of rare attack classes. The proposed system was trained and tested using the CSE-CIC-IDS2018 dataset, where the experimental results showed a high potential of the proposed CTGSM-DNN system in both binary and multiclass classification tasks. In our future work, we plan to leverage advanced techniques such as large language models (LLMs) and natural language processing (NLP) to further enhance the detection and classification of rare attacks particularly in scenarios where IoT devices generate complex or text-based data, such as logs or communication protocols.

\section*{Acknowledgment}
\setlength{\parindent}{0cm}
This work was sponsored by the 100 PhDs for Africa programme under the UM6P/EPFL Excellence in Africa Initiative.

\bibliographystyle{IEEEtran}
\bibliography{biblio_traps_dynamics}
\end{document}